\documentclass[apj1]{emulateapj}

\slugcomment{Received 2012 April 11; accepted 2013 December 28; published 2014 February 12}
\shorttitle{The MOND External Field Effect on the Dynamics of the Globular Clusters}
\shortauthors{Kamran Derakhshani}

\begin{document}

\title{The MOND External Field Effect on the Dynamics of the Globular\\ Clusters: General Considerations and Application to NGC 2419}

\author{Kamran Derakhshani}

\affil{Institute for Advanced Studies in Basic Sciences,
P. O. Box 45195-1159, Zanjan, Iran; kderakhshani@iasbs.ac.ir}

\begin{abstract}

In this paper, we investigate the external field effect in the context of the MOdified Newtonian Dynamics (MOND) on the surface brightness and velocity dispersion profiles of globular clusters (GCs). Using  N-MODY, which is an N-body simulation code with a MOND potential solver, we show that the general effect of the external field for diffuse clusters, which obey MOND in most of their parts, is that it pushes the dynamics towards the Newtonian  regime. On the other hand, for more compact clusters, which are essentially Newtonian in their inner parts, the external field is effective mainly in the outer parts of compact clusters. As a case study, we then choose the remote Galactic GC NGC 2419. By varying the cluster mass, half-light radius,  and mass-to-light ratio we aim to find a model that will reproduce the observational data most effectively, using  N-MODY.  We find that even if we take the Galactic external field into account, a Newtonian Plummer sphere represents the observational data better than MOND to an order of magnitude in terms of the total $\chi^2$ of surface brightness and velocity dispersion. \\

\end{abstract}

\section{Introduction}

The MOdified Newtonian Dynamics (MOND), proposed by Milgrom in 1983 (Milgrom 1983a) is one of the most serious rivals of the dark matter paradigm. It is a phenomenological theory which starts from a very simple ad hoc scaling assumption on the gravitational acceleration $g$. According to this assumption, for accelerations comparable to or less than a critical acceleration $a_0$, $g$ is greater than the Newtonian acceleration $g_N$ in a subtle way. This mild modification originally was innovated to explain the flat rotation  curves of spirals, and it proved to be highly successful to get this target (e.g. \citealp{san98}). The problems of the violation of energy and angular momentum conservation for non-spherical mass configurations in MOND \citep{fel84} were immediately fixed by presenting a Lagrangian formalism for it \citep{bek84}. Also, the covariant form of MOND was provided by Bekenstein who introduced a vector field, a scalar field and the gravitational tensor field, the so-called TeVeS \citep{bek04}, to explain relativistic effects such as gravitational lensing.

 Besides the rotation curves, this theory has had remarkable success in fitting to some other important observations of the local universe. However, MOND faces serious challenges on extragalactic scales. For example it can not completely explain the velocity of galaxies in clusters of galaxies (\citealp{ger92}; \citealp{san94}; \citealp{san99}; \citealp{agu01}). Also, the notorious offset between the baryonic and lensing masses in some galaxy cluster mergers \citep{clo06} has no convincing explanation by MOND (for a review see e.g., \citealp{fam12}).

As a matter of fact, MOND has been devised to substitute the assumption of dark matter. As Milgrom himself correctly recognized, his innovation would be in serious trouble in dynamical systems with no mass discrepancy but with internal accelerations $g\lesssim a_0$, e.g., open clusters \citep{mil83a}. To remedy this problem he designed his prescription in a way so that the Galactic  gravitational field could suppress the extra acceleration of MOND (the so-called External Field Effect, EFE). Thus, besides checking whether MOND can properly play the role of dark matter in, say, spiral galaxies, we can verify it as the correct dynamics in pure Newtonian systems. To do this, one should choose objects with low mass-to-light ratios and small internal accelerations comparable to $a_0$. Diffuse Galactic globular clusters (GCs) seem to be the best candidates \citep{bau05}. However, as we will elaborate in the next section, if the Galactic field is much higher than $a_0$ the internal dynamics will be Newtonian, regardless of the internal field. Therefore, distant Galactic GCs in the outer halo are the best candidates. Meanwhile, to study the EFE, which is absent in Newtonian dynamics, but is a trait of MOND (and every nonlinear dynamics), the gravitational field in which the GC is embedded should not be negligible. So a limited range with just a few Galactic GCs is at our disposal until the invention of telescopes and spectrographs with higher resolution powers will allow us to investigate kinematics of GCs in the halos of other neighbor galaxies.

 Several studies have endeavored  to discriminate between Newtonian dynamics and MOND using Galactic GCs (e.g., Jordi et al. 2009; Gentile et al. 2010; Baumgardt et al. 2005; Sollima et al. 2010, 2012; Haghi et al. 2009, 2011; Lane et al. 2009; Scarpa et al. 2011). However, until recently most of them used their overall (average) dynamical properties. With the development of high-resolution spectrographs, the velocity dispersion profiles of some GCs are curently available. Thus, more exquisite details of different models can be checked against the observational data.

 The GC NGC 2419 has recently received a great deal of attention in this regard since it has all the above-mentioned qualifications \citep{bau09}. It has also triggered a hot debate between the MONDian and Newtonian blocks (Ibata et al. 2011a, 2011b; Sanders 2012a, 2012b; for more details see Sect. 5). However, they all regarded this GC as isolated, i.e., with no external field. It is worth noting that Ibata et al. (2011a, hereafter I11a) deduced the ineffectiveness of the external field in alliviating the mismatch of MOND and the observational data of NGC 2419, using MONDian N-body simulation. Nevertheless, one should note that they performed this by applying the external field on the same best-fit models that they found in isolation.
 
In this paper we first investigate the effect of introducing the external field in the MONDian dynamics on the internal dynamics of GCs as a whole. Then we choose the Galactic GC NGC 2419 to check the EFE by finding the best-fit simulated model, having included the external field.

 In section 2 a brief exposition of MOND and the external field effect is presented. Section 3 describes the simulation approach  that is used to obtain dynamical predictions from the model. In section 4, a typical GC is modeled as a spherical collisionless system of stars with negligible binary fraction, and its internal dynamics is studied for a specific mass but different internal and external gravitational fields. Section 5 introduces NGC 2419 and fits a simulated  model to its data. Section 6 presents our conclusions and describes caveats one should be mindful of.

\section{MOND in a nutshell}

The popular interpretation of the MOND assumption is that the true gravitational acceleration $g$ is related to the Newtonian gravitational acceleration $g_N$ as the following \citep{mil83a}:

\begin{equation}
 g \mu(\frac{g}{a_{0}}) = g_{N}
\label{mond-1}
\end{equation}
where $a_{0}\approx1.2\times10^{-10} $ms$^{-2}$ is the characteristic acceleration of MOND, and $\mu(x)$ is a continuous monotonic function which is called the interpolating function and is characterized by the limiting conditions: $\mu(x)=x$ for $x\ll1$ and $\mu(x)=1$ for $x\gg1$. Though the major properties of the results in MOND are insensitive to the choice of $\mu(x)$ \citep{mil83b}, for detailed calculations a few forms of $\mu(x)$ have been proposed. The ''standard" form, $\mu(x)=x/\sqrt{1+x^2}$ \citep{mil83b}, and the ''simple" form, $\mu(x)=x/(1+x)$ \citep{fam05}, are the most well known. So, for $g\ll{a_0}$, we have $g=\sqrt{a_0g_N}$, and for $g\gg{a_0}$ the Newtonian gravitational field is recovered. For a more robust basis, MOND can be derived from a Lagrangian that yields the field equation in form of a modified Poisson's equation \citep{bek84}

\begin{equation}
 \nabla.[\mu(\frac{\parallel\nabla\phi\parallel}{a_{0}})\nabla\phi] = 4\pi G \rho
\label{modpois}
\end{equation}
where $\phi$ is the gravitational potential so that $\bf g = -\nabla\phi$, and $\parallel...\parallel$ is the Euclidean norm. In the case of high symmetries (spherical, cylindrical, plane) Eq.(\ref{mond-1}) will be concluded.\\
 Note that Eq.(\ref{modpois}) is a nonlinear equation. So if a system with internal acceleration ${\bf g}_{int} = -\nabla\phi_{int}$ is submerged in an external field $\phi_{ext}$   we can not simply substitute $\nabla\phi$ by $\nabla\phi_{int} + \nabla\phi_{ext}$ in general, unless the internal field can be considered a little perturbation in the external field. We should instead add $\rho_{ext}$, the source of the external field, to the right hand side and solve for the whole $\nabla\phi$. However, this would be prohibitively difficult to solve, for example for a distant GC moving in a gravitational field of the Galaxy. Nevertheless, many authors assumed  $\nabla\phi  \approx \nabla\phi_{int} + \nabla\phi_{ext}$ (\citealp{san02}, \citealp{zha06}, \citealp{wu07}, \citealp{wu08}, \citealp{kly09} and \citealp{hag11}) so that

\begin{equation}
 \nabla.[\mu(\frac{\parallel\nabla\phi_{int}+\nabla\phi_{ext}\parallel}{a_{0}})(\nabla\phi_{int}+\nabla\phi_{ext})]\approx 4\pi G \rho
\label{EFEmodpois}
\end{equation}

It is obvious that within the MOND framework any external gravitational field can have a drastic effect on the internal dynamics of self-gravitating systems. Nonlinearity of MOND causes the violation of  the strong equivalence principle (\citealp{mil83a}, 
\citealp{bek84}). Consequently, the internal dynamics of the system can be affected by the external field in which it is embedded, even in a uniform field. This is a characteristic trait of any nonlinear gravitation, unlike Newtonian dynamics which is linear and obeys the strong equivalence principle. This is why every Newtonian system embedded in a uniform field can be considered in isolation.

The asymptotic behaviors of this EFE can be categorized in three distinct regimes \citep{bau05}:

1) If  $g_{int}\gg{a_0}$ or $g_{ext}\gg{a_0}$, then the system is intrinsically Newtonian and its internal dynamics is independent of the external field (no EFE).

2) If  $g_{ext}\ll{a_0}$ and $g_{int}\ll{a_0}$, then MOND is completely satisfied (deep-MOND regime). This situation involves two special cases:

   2a) If  $g_{int}\ll{g_{ext}}\ll{a_0}$, then the system is in a quasi-Newtonian regime in which the dynamics is completely Newtonian, but with an effective gravitational constant $G=({a_0/g_{ext}})G_N$ where $G_N$ is the Newtonian gravitational constant.

   2b) If  $g_{ext}\ll{g_{int}}\ll{a_0}$, then the system is virtually isolated and its internal dynamics is MONDian, independent of the external field.

It is obvious that a sufficient change in the external field can cause a low-density self-gravitating system to switch between Newtonian, quasi-Newtonian, and isolated-MONDian dynamics. So the EFE, which is totally absent in Newtonian dynamics, is a decisive factor in discriminating between these two dynamics. Even if there was no need to presume the dark matter's existence or whether the dark matter will be eventually detected or not, checking for EFE would be essential in understanding the dynamics and evolution of low-field systems
\footnote{Milgrom himself called it "a most poweful test of the modified dynamics"\citep{mil83a}.}\citep{iba13}.

GCs are the best test beds for checking the above fact. Due to their low stellar mass-to-light ratios ($M_*/L\sim1$) (e.g., \citealp{mcl05}) they are commonly considered to be free of dark matter (e.g., \citealp{con11}). Most of them have low masses ($\sim 10^4-10^5 M_\odot$) resulting in accelerations less than or comparable to $a_0$. They are also highly spherical, so much simpler models can be used.

For an isolated GC in hydrostatic equilibrium we can numerically solve the Jeans equation along with Eq.(\ref{mond-1}) after assuming an appropriate model. This is the approach used by some authors(e.g., \citealp{san12a} and \citealp{iba11b}). 

If we take the same approach for a GC in a uniform external field we will have to use Eq.(\ref{EFEmodpois}) instead, which is too difficut to be solved even numerically. Many authors take a step further and use approximations such as $|{\bf g}_{int} +{\bf g}_{ext}| \approx g_{int} + g_{ext}$ (\citealp{gen07}; \citealp{fam07}; \citealp{hag09}) and  $|{\bf g}_{int} +{\bf g}_{ext}| \approx \sqrt{g_{int}^2 + g_{ext}^2}$ \citep{ang08}. This is certainly a course approximation that reduces a three-dimensional problem to a one-dimensional one, abandoning the general relative orientations of the acceleration vectors. 

On the other hand, if we do not use such approximations we will not be able to use the Jeans equation because introducing the external field will break the spherical symmetry. Therefore, we have to calculate the MONDian acceleration for each single particle.

\begin{figure}
\includegraphics[scale=0.7]{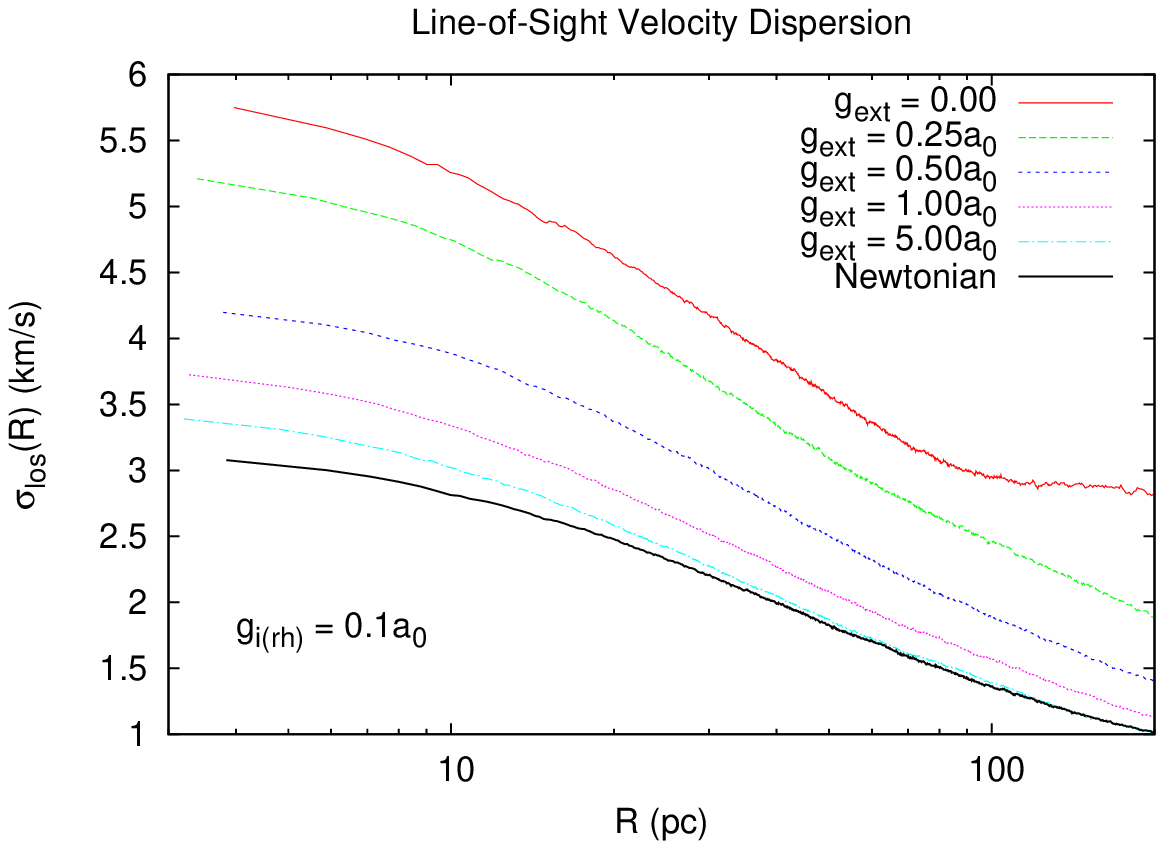}
\includegraphics[scale=0.7]{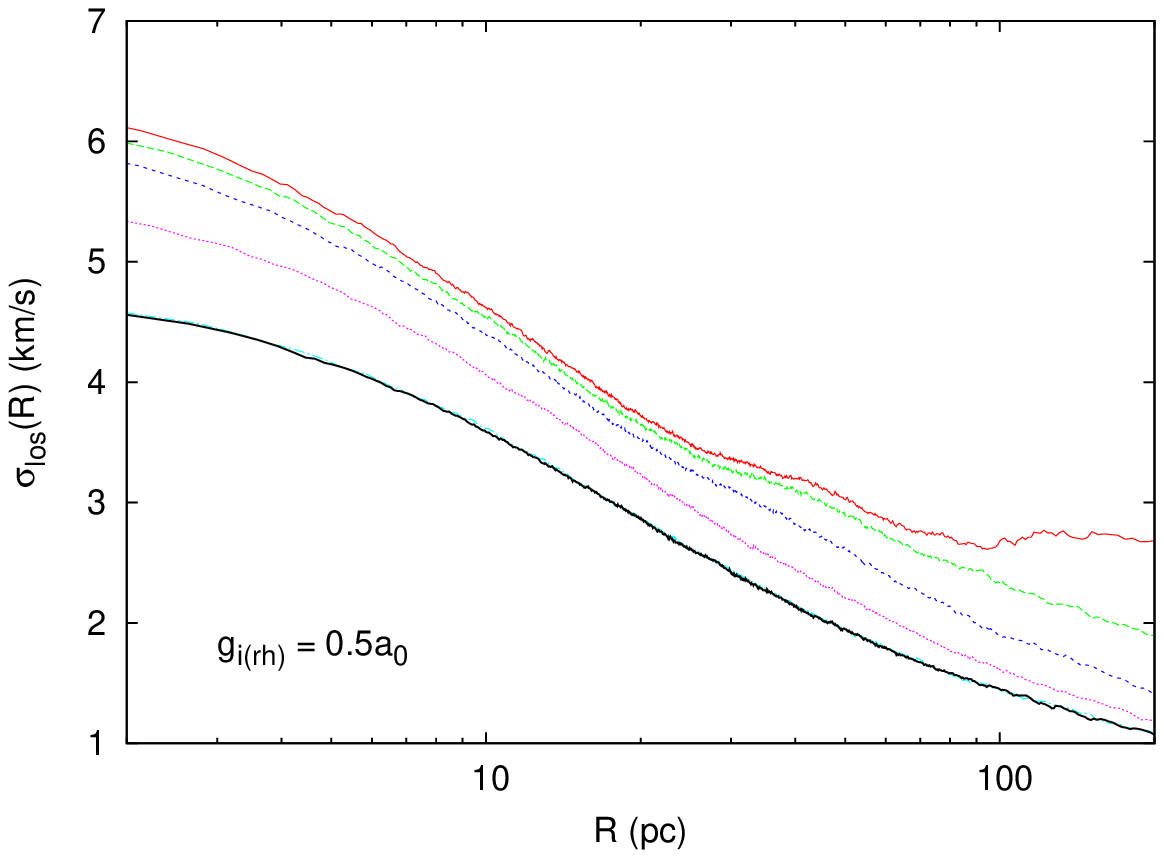}
\includegraphics[scale=0.7]{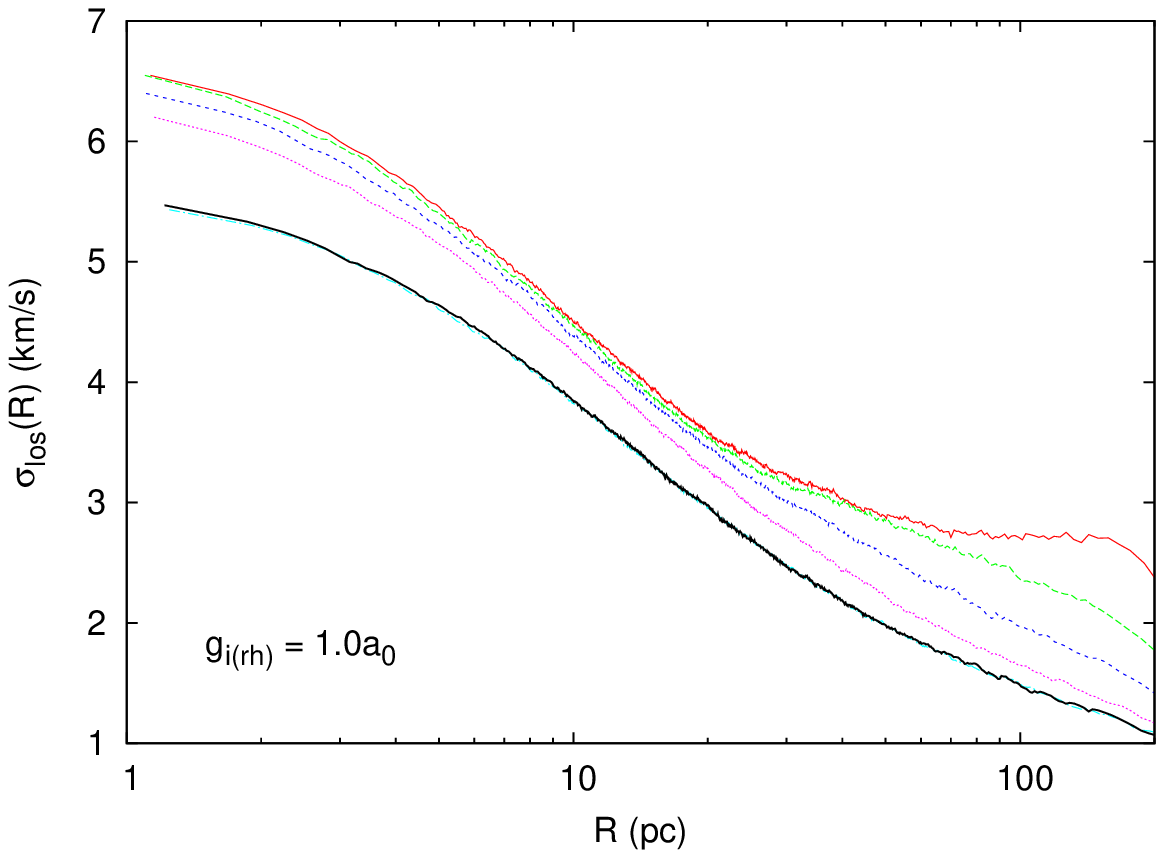}
\caption{Simulated line-of-sight velocity dispersion for a model globular cluster with a mass of $M=7.5\times10^5 M_{\odot}$, in Newtonian and MOND dynamics and in different internal and external gravitational fields.  The designated $g_{int}$ is the value of the internal field at the half-mass radius.}
\label{fig_1}
\end{figure}

\section{N-MODY Simulation}

We started from the modified Poisson equation

\begin{equation}
 \nabla.[ \mu(\frac{|{{\bf g}_{int} + {\bf g}_{ext}}|}{a_{0}}) ({{\bf g}_{int} + {\bf g}_{ext}})]\approx -4\pi G \rho_c
\label{final}
\end{equation}

subject to the boundary condition $-\nabla\phi \rightarrow {{\bf g}_{ext}}$ when $r \rightarrow \infty$ and where $\rho_c$ is the density of the cluster.

To simulate a system with this field equation we make use of the N-MODY code. This is a parallel particle-mesh code in three dimensions which has been developed and tested by Ciotti et al.(2006) and Nipoti et al.(2007) for the evolution of collisionless N-body systems in either MONDian or Newtonian dynamics. It uses a grid in spherical coordinates with $N_r\times N_\theta\times N_\phi$ cells for the leap-frog time integration. The current version of N-MODY uses the standard interpolating function.

N-MODY was originally designed for isolated systems. To introduce a constant external field we employ the same changes which have already been made by Haghi et. al.(2011).  A constant external acceleration is added vectorially to the internal acceleration of each particle in each step. This method has proven to agree very well with the asymptotic regimes mentionded in section 2 \citep{hag11}. 

As the initial equilibrium condition, we use the Newtonian Plummer model \citep{plu11}, which is characterized by two parameters; the total stellar mass $M$ and the half-light radius $r_h$.

We extract the desired quantities, densiy $\rho(r)$ and radial velocity dispersion $\sigma_r(r)$-averaged over bins of 100 stars, from our simulation's outputs. To compute the observational quantities, we have the following relations \citep{bin10}

\begin{equation}
\Sigma(R) = 2\int^\infty_R\frac{r\rho(r)}{\sqrt{r^2 - R^2}}dr
\end{equation}
and
\begin{equation}
\Sigma(R)\sigma_{los}^2(R) = 2\int^\infty_R\rho(r)\sigma_r^2(r)\frac{rdr}{\sqrt{r^2 - R^2}}
\end{equation}

where $R$ is the projected radius, $\Sigma(R)$ is the surface mass density, and $\sigma_{los}(R)$ is the line-of-sight velocity dispersion. In order to relate $\Sigma(R)$ to the observational quantity $\mu_V(R)$ (the surface brightness in terms of $mag.arcsec^{-2}$), we use the standard relation $\mu_V(R) = -2.5log I(R) + 26.422$ \citep{lan99} in which $I(R)$ is the luminosity surface density in terms of $L_{\odot}pc^{-2}$. Assuming a reasonable $M_*/L$ we will directly have $I(R)$.

\begin{figure}
\includegraphics[scale=0.7]{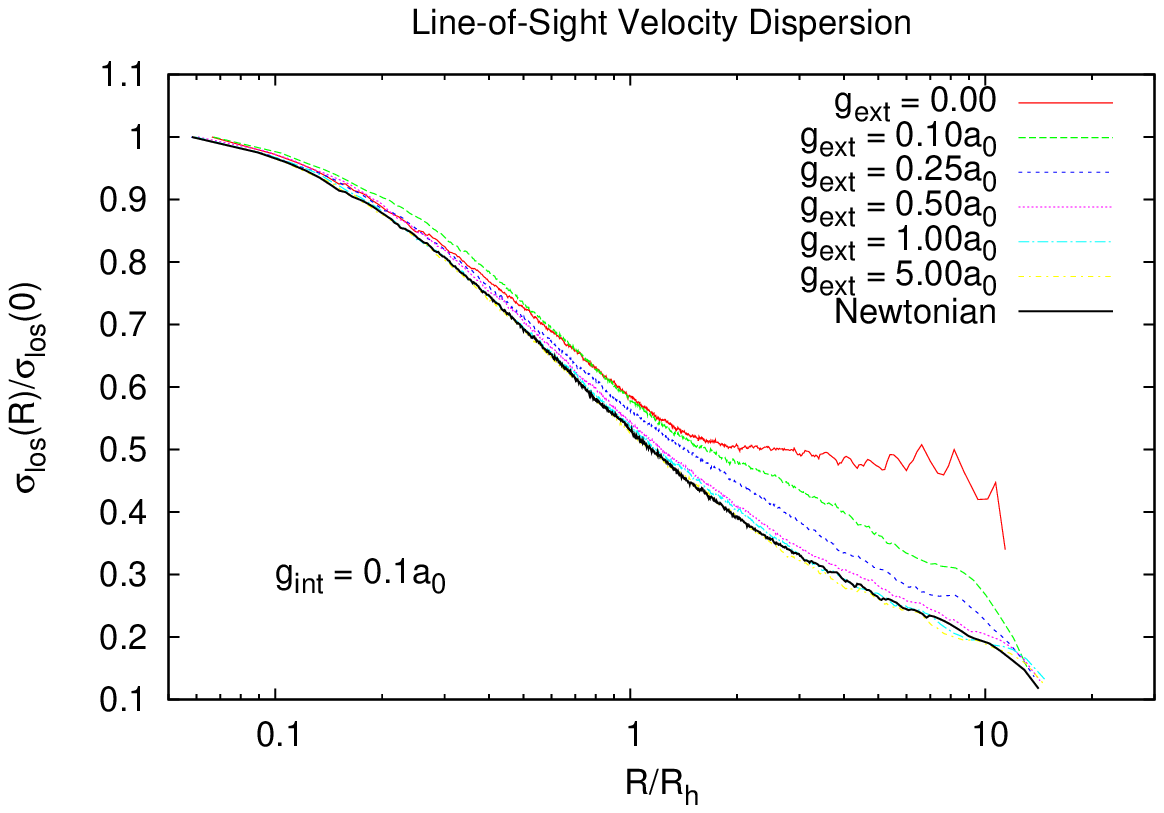}
\includegraphics[scale=0.7]{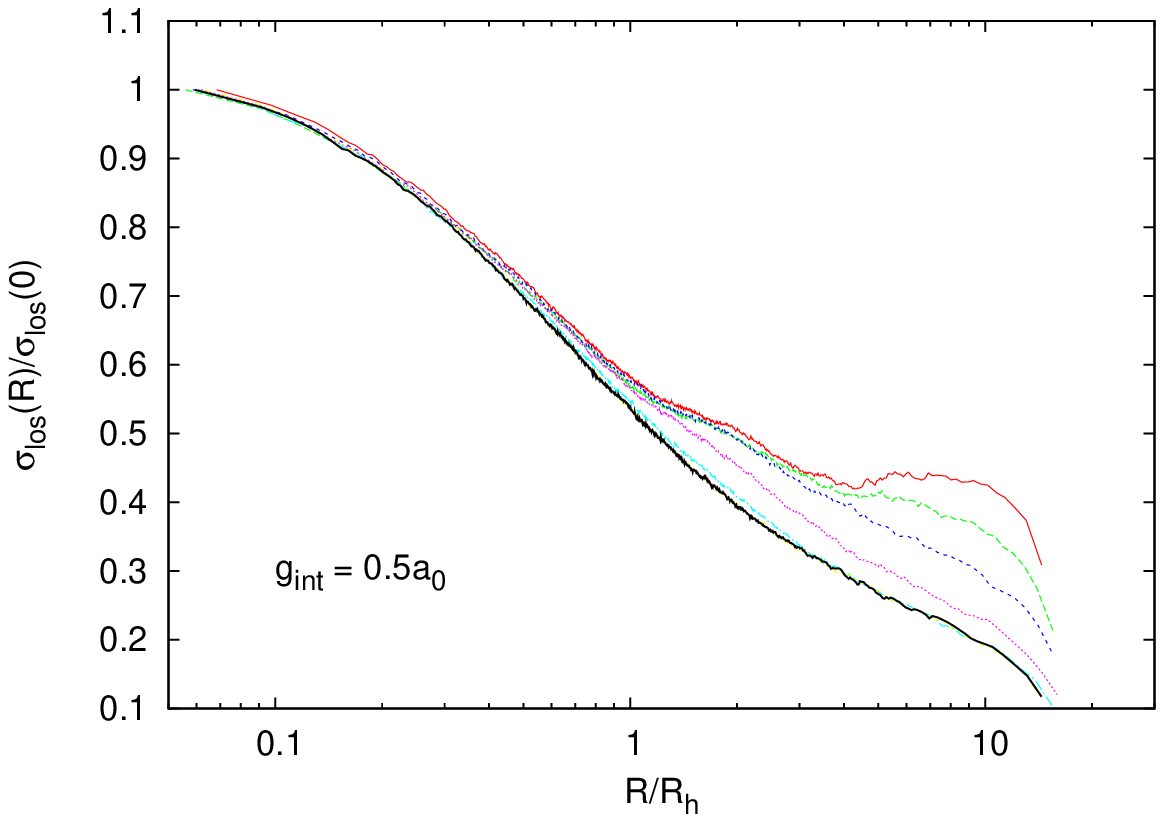}
\includegraphics[scale=0.7]{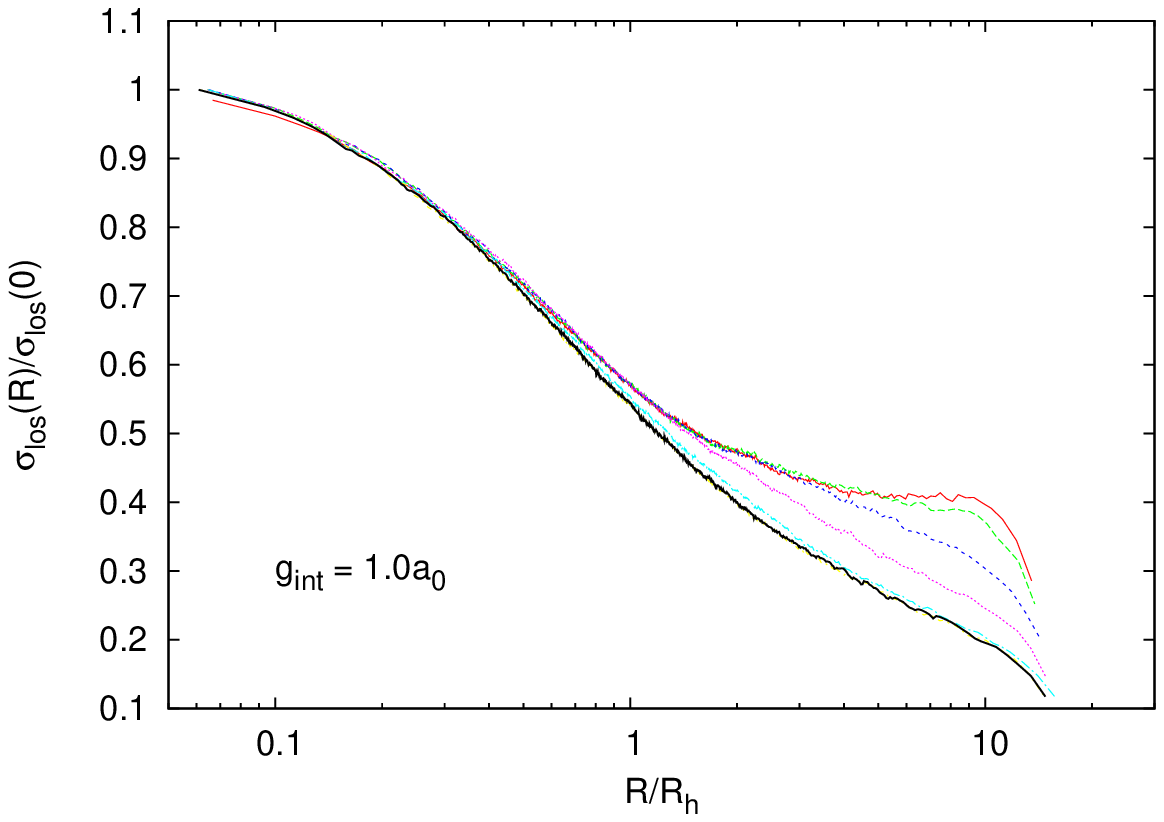}
\caption{Same as Figure 1, normalized to the half-mass radius $R_h$ and central velocity dispersion $\sigma_{los}(0)$.}
\label{fig_2}
\end{figure}

\begin{figure}
\includegraphics[scale=0.7]{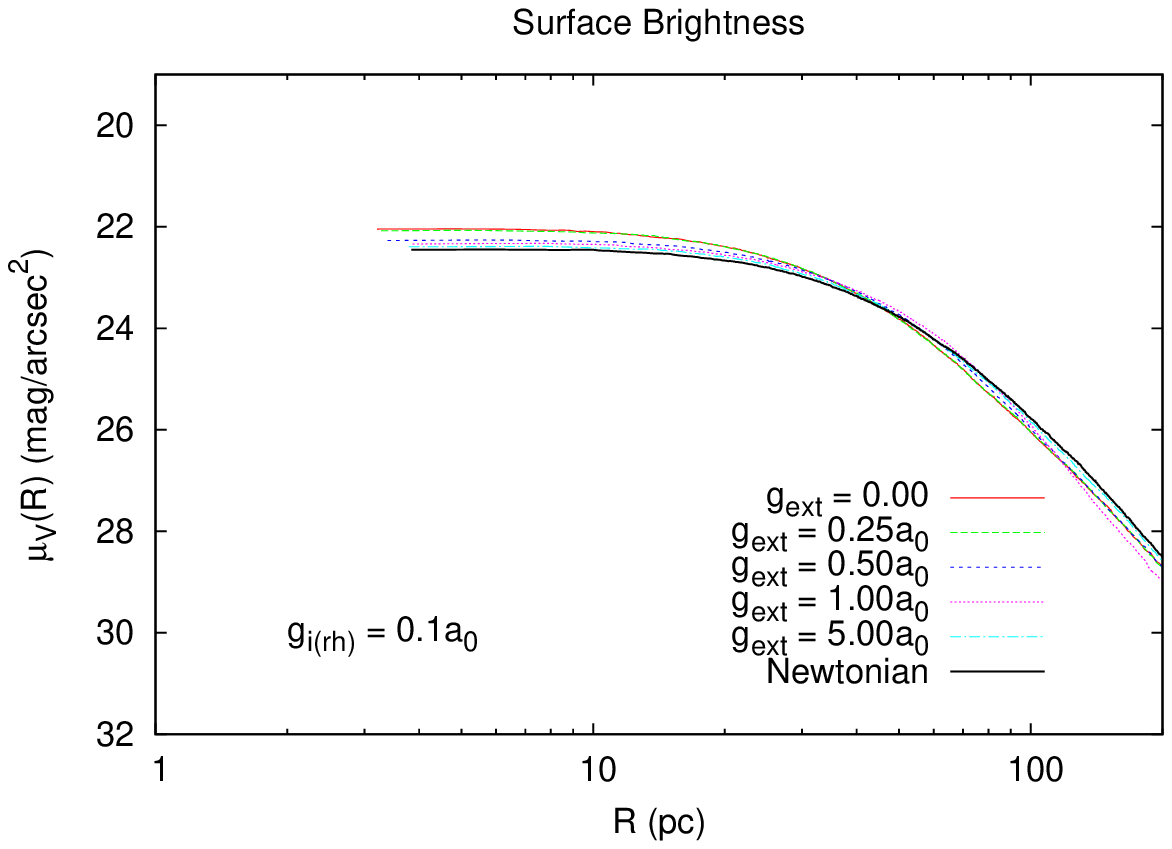}
\includegraphics[scale=0.7]{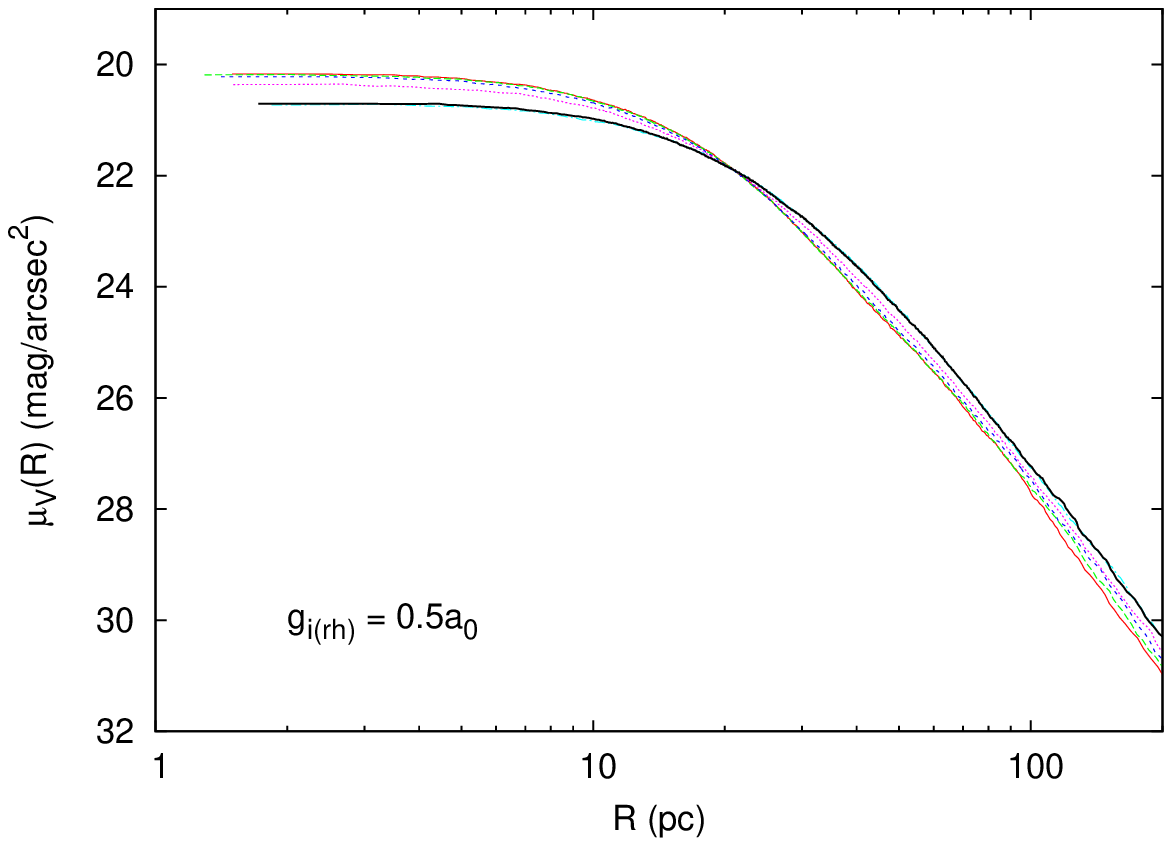}
\includegraphics[scale=0.7]{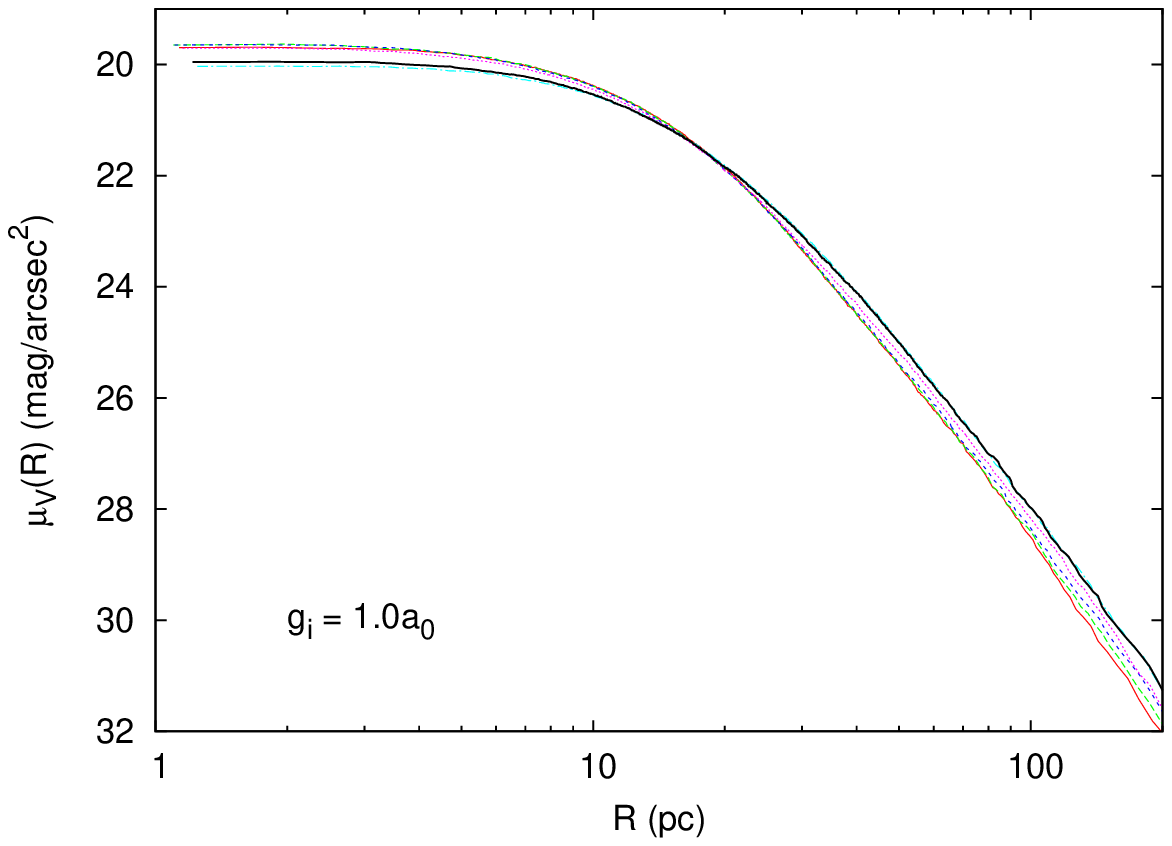}
\caption{Simulated surface brightness for a model globular cluster with a mass of $M=7.5\times10^5 M_{\odot}$ and a stellar mass-to-light ratio $M_*/L=2.0$, in Newtonian and MOND dynamics and in different internal and external gravitational fields. The designated $g_{int}$ is the value of the internal field at the half-mass radius.}
\label{fig_3}
\end{figure}

\section{Simulations of GCs: The General Case}

To study the effect of the external field on the internal dynamics of a typical GC, we model it as a spherical system of $N=10^5$ stars with a mass $M = 7.5\times10^5 M_{\odot}$ for different amounts of $g_{int}$  and $g_{ext}$. The chosen values - in terms of $a_0$ - are 0.1, 0.25, 0.5, and 1 for the $g_{int}$, and 0, 0.1, 0.25, 0.5, 1.0, and 5 for the $g_{ext}$. As a criterion for the $g_{int}$ we designate its value at the half-light radius $r_h$, i.e., $g_{int}(r_h) = GM/2r_h^2$. The initial equilibrium condition we use is the Plummer model in the Newtonian regime which is characterized by the total mass and the Plummer radius. The Plummer radius is approximately equal to $r_h/1.3$. So choosing a value for $g_{int}(r_h)$ fixes the Plummer model. N-MODY evolves this model under a MOND potential and after a few dynamical times we will have a GC in MONDian equilibrium. An excerpt of the resulted velocity dispersion and surface brightness profiles is shown in Figures 1-3.

\begin{figure}
\includegraphics[scale=0.7]{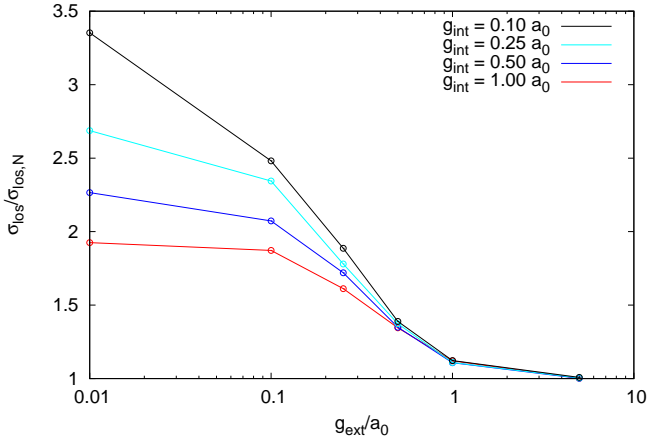}
\caption{Relative overall line-of-sight velocity dispersion of the simulated model globular cluster with a mass of $M=7.5\times10^5 M_{\odot}$ and a stellar mass-to-light ratio $M_*/L=2.0$, as a function of the Galactic external field and for various internal fields. $\sigma_{los,N}$ is the Newtonian overall line-of-sight velocity dispersion.}
\label{fig_4}
\end{figure}

A glimpse shows that there is a remarkable difference between MOND and Newtonian dynamics in GCs. Figures 1 and 2 show that in all conditions the MONDian velocity dispersion is over the Newtonian one.  Also, it can be seen that as the external field gets stronger the cluster dynamics tends towards the Newtonian prediction. In particular, it  is obvious that even a small external field erases the flat behavior of the velocity dispersion profile at large radii. For more compact GCs (higher internal fields), this effect is remarkable mostly in the outer parts.

 These findings are verified in the obtained profiles for the surface brightness (Figure 3). MOND increases the surface brightness of an isolated GC, especially in the outer parts. Again, the effect of the galactic field is weakening this effect and pushing the profile toward the Newtonian profile. Also it can be inferred that for stronger internal fields the differences are decreased.

To investigate the EFE on the internal dynamics of GCs in terms of overall velocity dispersions, we calculated the  overall Newtonian, isolated MOND, and MOND with external field line-of-sight velocity dispersions ($\sigma_{los,N}$, $\sigma_{los,M}$, and $\sigma_{los}$, respectively)  and figured the ratios $\sigma_{los}/\sigma_{los,N}$ and $(\sigma_{los} - \sigma_{los,N})/(\sigma_{los,M} - \sigma_{los,N}$), as functions of the external field for different internal accelerations (Figures 4 and 5). In accord with the predicted asymptotic behaviors mentioned in Section 2, Figure 4 shows that for large $g_{int}$ or large $g_{ext}$ the overall $\sigma_{los}$ approaches a Newtonian flat curve, i.e., independent of the external field, though with different asymptotic velocity dispersions. It can also be seen that diffuse GCs with $g_{int}\ll a_0$ are much more sensitive to the changes of the external field.

 Figure 5 shows the relative distances of the isolated MOND and MOND with an external field, from the Newtonian regime in each case. We can see that for vanishing external fields they have the same differences from the Newtonian values, as expected, while for large external fields the ratio approaches zero, meaning that the galactic field forces the kinematics to be Newtonian.\\

 \begin{figure}
\includegraphics[scale=0.7]{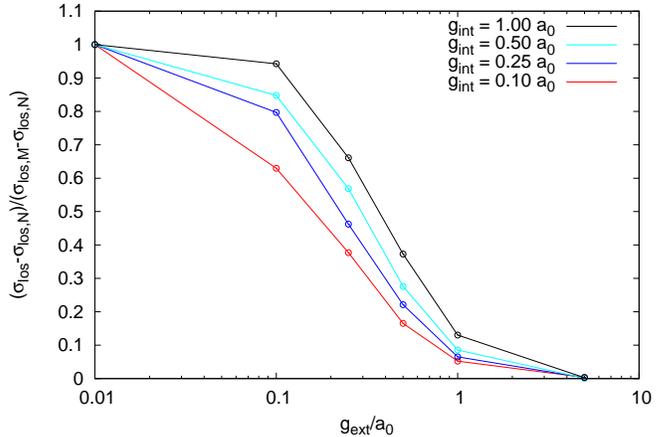}
\caption{Ratio of difference overall line-of sight velocity dispersions of the simulated model globular cluster with a mass of $M=7.5\times10^5 M_{\odot}$ and a stellar mass-to-light ratio $M_*/L=2.0$, as a function of the galactic external field and for various internal fields. $\sigma_{los,N}$, $\sigma_{los,M}$, and $\sigma_{los}$ are the Newtonian, isolated MOND, and MOND with an external field overall line-of-sight velocity dispersions, respectively.}
\label{fig_5}
\end{figure}

\begin{table*}[htbp]
\caption{The best-fit simulated models for NGC 2419 and their goodnesses of fit.}
\label{tab2}

\begin{center}
\begin{tabular}{cccccccc}
\hline
Model&$M$&$r_h$&$M_*/L$&$\mu_V(0)$&$\sigma_{los}(0)$&$\rho(0)$&$\chi_{total}^2$\\
&$(10^5M_{\odot})$&$(pc)$&&$(mag/arcsec^2)$&$(km/s)$&$(M_{\odot}/pc^3)$&\\
\hline
Newtonian Plummer&$9.0$&$25$&$1.9$&$20.24$&$6.53$&$26.68$&$48$\\
MOND+EFE&$8.0$&$25$&$1.8$&$20.27$&$6.25$&$25.10$&$119$\\
MOND-isolated&$8.0$&$26$&$1.9$&$20.43$&$6.24$&$16.47$&$126$\\
\hline

\end{tabular}
\end{center}

\tablecomments{Newtonian Plummer: The best-fit Plummer sphere in the Newtonian equilibrium; MOND+EFE: The best-fit model in the MONDian equilibrium including a Galactic external field $0.14a_0$; MOND-isolated: The best-fit model in the MONDian equilibrium in isolation. The fit parameters are the mass $M$, the half-light radius $r_h$, and the stellar mass-to-light ratio $M_*/L$. The derived parameters are central surface brightness $\mu_V(0)$, central line-of-sight velocity dispersion $\sigma_{los}(0)$, and the central mass density $\rho(0)$. The $\chi_{total}^2$ is the minimized sum of $\chi_{\mu_V}^2$ and $\chi_{\sigma_{los}}^2$.}

\end{table*}

\section{A case study: NGC 2419}
NGC 2419 satisfies all the qualifications mentioned in Section 1. Its Galactocentric radius is $R_{GC} \approx 87.5\pm3.3$ kpc \citep{dic11}, its  half-light-radius is $r_h \approx$  23pc, on the basis of Harris (1996), and its tidal radius is $r_t \approx$ 350pc \citep{rip07}. Meanwhile, with ellipticity $\approx 0.03$ it can be assumed highly spherical  \citep{har96}. Moreover, although it is relatively massive ($9.0\pm2.2\times10^5 M_{\odot}$; Baumgardt et al. 2009), which means it has a  sufficient number of stars and large velocity dispersions for reliable statistical analysis, its half-light radius is among the largest of the Milky Way GCs, i.e., it is not compact. Thus, we expect its internal gravitation to be mostly at the order of or less than $a_0$ so that a notable fraction of cluster stars lie within the MOND regime. Given the large distance and the relatively large mass of the cluster, any tidal effect (both in terms of shocks and virial perturbations) is expected to be negligible. These factors have caused several authors to pay attention to this GC in recent years (Baumgardt et al. 2009; Sollima et al. 2010; Sanders 2012a; I11a).

Recently, I11a solved the Poisson and the modified Poisson equations for a very large set of parameter settings for the King and Michie models which were sampled through a Markov-Chain Monte Carlo method. They selected NGC 2419 as a ''crucible'' to settle the issue between the two gravitation theories. They contrasted their results to high-resolution data and, based on a likelihood analysis, came to the conclusion that the best Newtonian Michie model fits to the observations with a $\sim$ 40,000 likelihood over the best MONDian Michie model, a "very severe challenge for MOND". They also claimed that by using the MOND N-body simulation (N-MODY code) and the same density function obtained from their best-fit model, they accounted for the EFE for two different $g_{ext}=0.1a_0$ and $0.2 a_0$ and they could not salvage MOND from this challenge.

 In response, Sanders  \citep{san12a} reported that by using a polytropic model for the GC, a good fit to the same data is possible. Therefore, NGC 2419 could not be used as a MOND-violating object. Then, Ibata and colleagues tried to find the best polytropic model in MOND gravity and contrasted it to the best Michie model in Newtonian gravity they had found for NGC 2419. They claimed that the latter is about 5000 times more likely than the former, hence they rejected MOND \citep{iba11b}.

 In sequel, Sanders claimed that a polytropic model with "running polytropic index" could lead to an improved representation of the observed data of NGC 2419 \citep{san12b}.  

However, we suspect that introducing the Galactic external field may have a remarkable effect on the internal dynamics of NGC 2419 in favor of MOND. Although the Galactocentric distance of this GC is very large, so that the Galactic external field is just about $0.14a_0$, the internal acceleration is well below the $g_{ext}$ in its outer parts. Hence we conjecture that the surface brightness and velocity dispersion profiles of this GC, as a MONDian sphere, may be affected by the $g_{ext}$, at least in the outer parts.

Here we compare surface brightness and velocity dispersion profiles under two contexts: Newtonian and MONDian with external field.

\begin{figure}
\includegraphics[scale=0.7]{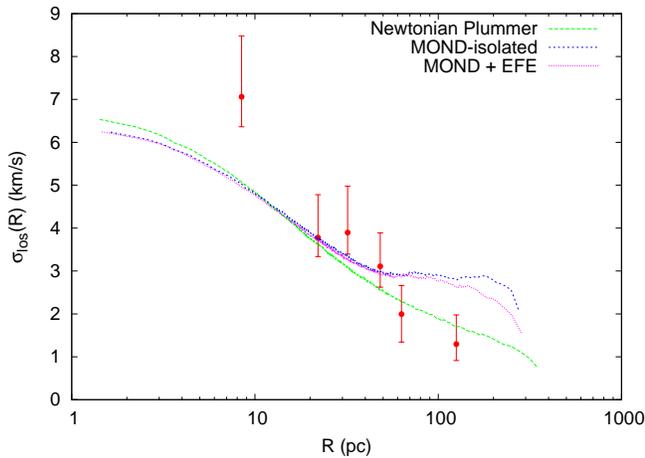}
\caption{Observed line-of-sight velocity dispersion of NGC 2419 against the best-fit simulated models. The values of $\chi^2_{\sigma_{los}}$ for the Newtonian Plummer, MOND+EFE, and MOND-isolated are $2.0$, $4.2$,and $4.7$, respectively (For the nomenclature and best-fit parameters refer to the caption of Table 1)}
\label{fig_6}
\end{figure}

\begin{figure}
\includegraphics[scale=0.7]{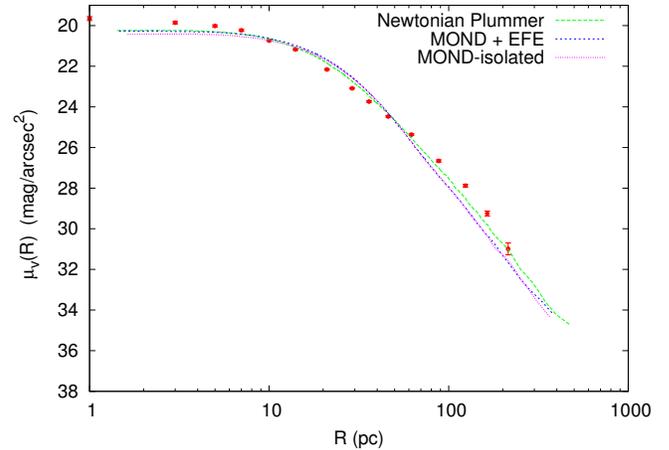}
\caption{ Observed surface brightness of NGC 2419 against the best-fit simulated models. The values of $\chi^2_{\mu_V}$ for the Newtonian Plummer, MOND+EFE, and MOND-isolated are $46.2$, $114.8$, and $121.9$, respectively (For the nomenclature and best-fit parameters refer to the caption of Table 1)}
\label{fig_7}
\end{figure}

\begin{figure}
\includegraphics[scale=0.7]{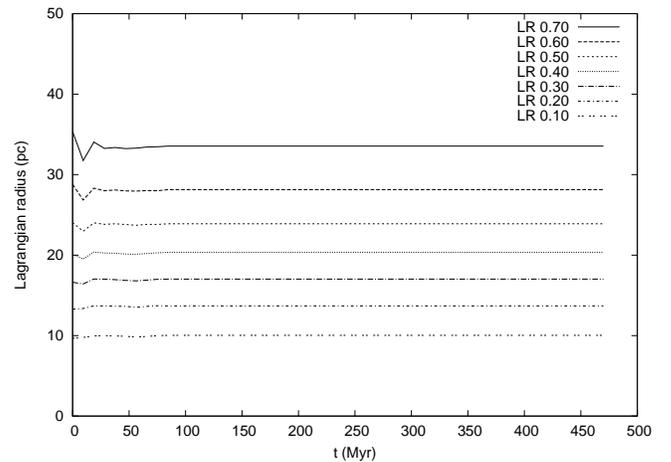}
\caption{Lagrangian radii of the Newtonian Plummer sphere evolving to the best  MONDian model embedded in the Galactic external field of $0.14a_0$. LR10, LR20,... are the radii of the spheres containing 10\%, 20\%,... of the total cluster mass, respectively.}
\label{fig_8}
\end{figure}

To investigate the EFE on NGC 2419, we simulate a cluster consisting of $N=10^5$ stars with positions and velocities chosen according to a Plummer profile with a broad  range of masses from $10^5$ to $10^6 M_{\odot}$ and half-light radii from 18 to 27 parsecs, including the $r_h=$ 23pc estimated by Harris (1996).  The external field is calculated as $g_{ext} = V_c^2/R_{GC} \approx 0.14a_0$ , where $V_c= 220$ km/s is the asymptotic rotational velocity of the Galaxy. Fitting of the calculated velocity dispersion and surface brightness profiles to the observed data points is achieved by adjusting the cluster mass $M$, the half-light radius $r_h$, and the stellar mass-to-light ratio $M_*/L$, as three free parameters of our analysis.  We assume a constant $M_*/L$ throughout the cluster and a range of $1.0-3.0$ that embraces the range $1.2-1.7$ recently obtained by Bellazzini et al. (2012).

We use $\chi^2$ to compare the observed velocity dispersion and surface brightness profiles to those of the simulated models.

In order to have a single model that matches the photometric and dynamical data simultaneously, we minimized the sum of the two $\chi^2$s. In doing so, we found that for $M= 9.0\times 10^5 M_{\odot}$, $r_h=25$ pc,  and $M_*/L=1.9$ the Newtonian Plummer model gives the minimum $\chi_{tot}^2 \approx 48$ while for the MONDian model-including the external field-the minimum $\chi_{tot}^2 \approx 119$ for the mass $M= 8.0\times 10^5 M_{\odot}$, $r_h=25$ pc, and $M_*/L= 1.8$ (Figures 6 and 7).

Overall results are summarized in Table 1. For comparison, the corresponding results for the isolated MONDian model are reported, too. Clearly the Newtonian dynamics can yield remarkably better fits to the observational data of NGC 2419 than MOND, even if we take the external field into account. The best-fit model we found for the system (NGC 2419) in MOND in an external field is  $\sim10^2$ times less likely than the best-fit Newtonian model we found for the same system.

The Plummer sphere that we used as the initial condition in our simulation is intrinsically isotropic. To ensure that our model was stable we plotted the evolution of the Lagrangian radii up to  a few crossing times. As shown in Figure 8, after a rapid collapse the system evolves for several crossing times to reach the equilibrium state and its half-mass radius remains nearly constant for the rest of the cluster evolution.
\\

\section{Conclusions and caveats}

In this paper we studied the velocity dispersion and surface brightness of simulated GCs in both strong and weak external fields in MOND. We showed that the MOND dynamics differs greatly from the Newtonian one in GCs. In general, the velocity dispersion of a typical GC in MOND is larger than that predicted by the Newtonian dynamics, especially in the outer regions. A uniform external field causes the internal kinematics to drift toward the Newtonian regime, overwriting the flattening of the velocity dispersion profile at large radii \footnote{Note that such a flat velocity dispersion profile is expected only for models without tidal truncation like the Plummer models adopted here.}. These effects are more spectacular in diffuse GCs (where $g_{int}\ll a_0$), while in compact GCs they are apparent only in the outer regions. \\
In terms of the overall velocity dispersion, the simulated dynamics is in harmony with MOND and its asymptotic behaviors (see Section 2). For every fixed external field, increasing the internal field causes the velocity dispersion to approach the Newtonian value. On the other hand, the same occurs if we fix the internal field and increase the external field.

Nevertheless, to explain the internal dynamics of  the Galactic GC NGC 2419, introducing the Galactic external field in the model may not be in MOND's favor. In terms of $\chi^2$ goodness-of-fit, our best Newtonian Plummer model is better than the best model in MOND with an external field to an order of magnitude. So this disproves the effectiveness of EFE in MOND in matching it to the observations. Our analysis showed that the main change occurs in the outer parts of the GC, as might have been conjectured, because the external field can show up itself where the star accelerations are much smaller than $a_0$. Meanwhile, the mass of the best Newtonian model ($9.0\times10^5 M_{\odot}$) and its half-light radius ($25$ pc) are comparable with $9.12\times10^5 M_{\odot}$ (obtained by I11a) and $23$ pc (estimated by Harris 1996), respectively, and its $M_*/L$ ($1.9$) is marginally in accord with the constraints found recently by Bellazzini et al.(2012). Therefore, NGC2419 is similar to the other outer Galactic halo clusters  Pal 14 \citep{jor09} and Pal 4 \citep{fra12} in that it is difficult to explain their internal dynamics with MOND. 

However, we should keep some caveats in mind. The Plummer model may not be the best one to describe NGC 2419. Other suitable models might be checked. The chosen interpolating function $\mu(x)$ may not be the best choice. Other forms of interpolating functions may yield different results.

Our assumptions should be considered as well. The most critical being that in deriving the surface brightness we assumed a constant mass-to-light ratio, while in general  this is not strictly the case, because of the mass segregation in star clusters (Frank et al. 2012; Jordi et al. 2009). So the plausibility of this assumption is debatable and a radially-varying $M_*/L$ with a suitable profile might conclude in a better fit. However, specifically for NGC 2419, I11a argue that the lack of any significant mass segregation, proposed by Dalessandro et al.(2008) on the basis of observational results  on the radial distribution of blue
straggler stars, shows that the mass-to-light ratio can be reasonably assumed to be constant with radius. Another implicit assumption in our treatment is that we considered ${\bf g}_{ext}$ at the position of NGC 2419  constant in time, equivalent to the assumption of a circular orbit. As a matter of fact, the orbit of this GC is unknown and in the case of an elliptical orbit the Galactic field changes with time and this affects the internal dynamics. However, I11a estimated that the internal dynamical time of this GC is much less than its orbital period. So the assumption of a constant external field equal to the present value is plausible. The effects of non-sphericity and some rotation should also be evaluated.

As a GC, NGC 2419 may include some fraction of binaries. Unless the binary star fraction in NGC 2419 is inappreciable, it can inflate the observed velocity dispersion (Cote et al. 2002).  However, I11a showed that although the fraction of binaries  in NGC 2419 is estimated to be as large as 20\%, their velocity distribution is so peaked around $v=0$ that their effects on the velocity dispersion of NGC 2419 is negligible. At the largest amount, one can invoke the uncertainties on the binary parameters such as mass ratios, orbital eccentricities, period distributions, and mass functions to evaluate their impact on the velocity dispersion.

In regards to the EFE, like many other authors, we assumed that the total MONDian acceleration would equal the vectorial sum of internal and external accelerations. However, abandoning this assumption would require a high-resolution MOND simulation with the ability to embrace both Galaxy and cluster distributions to solve the modified Poisson equation. Such a simulation is not available at this time. 

The assumption of isotropy in the simulation code which is used in this study  (N-MODY) might cast a shadow on the results. Certainly not all elliptical structures can be assumed isotropic. In fact, van Albada showed that the formation of stellar systems through dissipationless gravitational collapse leads to isotropic cores and radially-anisotropic envelopes \citep{alb83}. Nowadays, anisotropy attracts more attentions in galactic astrophysics. As an example, a by-product of I11a is that the best Newtonian Michie model is more likely to describe NGC 2419 than the best Newtonian King model by a factor of $10^{118}$. Some changes have to be made in the model-producing programs to include various models of anisotropy, the next step.

\end{document}